# The amplitude of the initial density fluctuation spectrum from lensing statistics

Eelco van Kampen

*Royal Observatory Edinburgh, Blackford Hill, Edinburgh EH9 3HJ, U.K.*

**Abstract.**
A complete catalogue of galaxy cluster models can be used to constrain cosmological parameters like $\sigma_8$, the amplitude of the initial density fluctuation spectrum. We focus here on two strong lensing statistics for clusters of galaxies: the fraction of rich Abell clusters that are critical to lensing, and the abundance of arcs in such clusters, especially the X-ray luminous ones. The specific aim of this paper is to constrain $\sigma_8$ for the standard Cold Dark Matter scenario within an Einstein-de Sitter universe.

## 1. Introduction

A rich cluster of galaxies can, if sufficiently massive, produce *arcs*, distorted and magnified images of background galaxies. In this short paper we try to exploit the fact that the occurance of arcs depends on the cluster mass. In most cosmological scenarios cluster masses increase with time, so the abundance of arcs should increase with time as well.

We study this for the CDM $\Omega_0 = 1$ scenario, with the specific goal of constraining the amplitude of the initial density fluctuation spectrum, as quantified by the parameter $\sigma_8$, the *r.m.s.* fluctuation of the 8 $h^{-1}$Mpc Top-Hat smoothed density field. This parameter is fixed once the cosmological timing is fixed. For this purpose a complete sample of cluster models is used, whose arc statistics are matched to an observational arc survey to constrain $\sigma_8$.

## 2. The model cluster catalogue

We devised a technique for the construction of a model catalogue in order to construct a fair sample of high-resolution models of galaxy clusters (van Kampen 1994). This technique is an alternative to extracting clusters from a single low-resolution simulation of large-scale structure formation, as has been done by eg. Frenk et al. (1990). The actual catalogue that we produced was constructed to mimic an observed sample (Mazure et al. 1995; Katgert et al. 1995).

Individual cluster models were build using a dissipationless N-body code which was supplemented with a recipe for galaxy formation and merging (van Kampen 1994). This recipe is necessary in order to retain any galaxies at all in the final cluster. Groups of particles representing galaxies in a standard dissipationless N-body run get disrupted due to two-body heating (Carlberg 1994; van Kampen 1994, 1995).

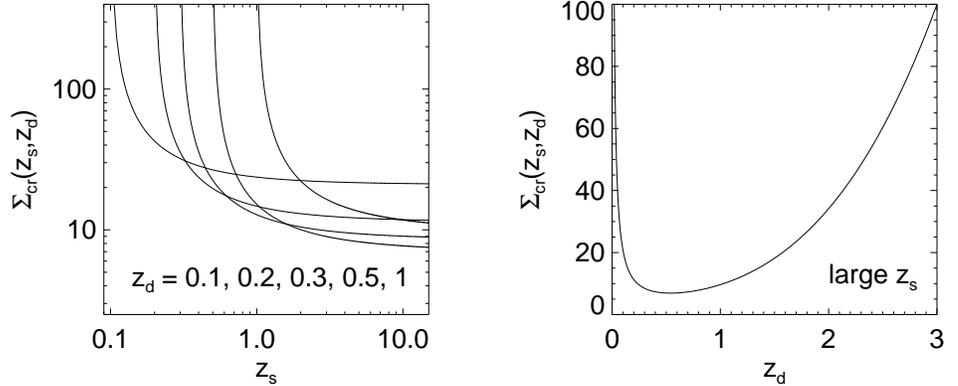

Figure 1. Critical surface mass density as a function of the lensing cluster redshift $z_d$ and the source redshift $z_s$ for an Einstein-de Sitter universe. The units of $\Sigma_{cr}$ are arbitrary.

## 3. Cluster statistics and $\sigma_8$

Using our simulation technique, i.e. including a recipe for galaxy formation, we found that the observed galaxy autocorrelation function sets $\sigma_8$ to the range $0.45 - 0.55$ for the standard CDM scenario (van Kampen 1994). Furthermore, we used the simulation technique to construct a complete catalogue of cluster models. The richness distribution of these clusters is also a monotonically growing function of time. Comparing it to the observed richness distribution we found that $\sigma_8 \approx 0.5$ (van Kampen 1994). We also found there that the cumulative distribution of line-of-sight velocity dispersions obtained for an observed complete sample (Mazure et al. 1995; Katgert et al. 1995) fits the model distribution for $\sigma_8 \approx 0.4$ rather well. Thus we have a relatively consistent (but not perfect) determination of the initial amplitude of the standard CDM spectrum from these three measures on cluster scales, which are likely to be relatively independent.

## 4. Arc statistics and $\sigma_8$

### 4.1. Fraction of clusters critical to lensing

A cluster will only be capable of producing arcs if its surface mass density is larger than a *critical* surface density $\Sigma_{cr}$, which is a function of the chosen cosmology, the redshift $z_d$ of the lensing cluster, and the redshift $z_s$ of the source

(eg. Schneider, Ehlers & Falco 1992). For an Einstein-de Sitter universe we have

$$\Sigma_{\rm cr} = \rho_{\rm cr}\frac{c}{3H_0}\frac{(1+z_{\rm d})^2(\sqrt{1+z_{\rm s}}-1)}{(\sqrt{1+z_{\rm d}}-1)(\sqrt{1+z_{\rm s}}-\sqrt{1+z_{\rm d}})} \ .$$

This function is small for large $z_{\rm s}$ and $0.2 < z_{\rm d} < 1$, as can be seen from Figure 1. The average redshift for rich Abell clusters in this redshift range is approximately 0.35. For this $z_{\rm d}$ the average source redshift $z_{\rm s}$ is about one. Using these optimal values we can calculate the fraction of rich Abell clusters within the $z_{\rm d} \approx 0.35$ redshift shell that are critical to lensing as a function of $\sigma_8$, and therefore the *maximal* fraction for any Abell cluster sample. The dependency on $\sigma_8$ is clearly demonstrated in Figure 2, where we plot surface mass densities for cluster model 2 from the catalogue (see van Kampen 1994 for its properties) positioned at $z_{\rm d} = 0.35$, and overlay these with the critical lines for a source at $z_{\rm s} = 1$. After analysing all model clusters from our catalogue, we obtained a single fraction of critical rich Abell clusters for each value of $\sigma_8$. The results are listed in Table 1.

Table 1. Fraction of rich Abell clusters critical to lensing at $z_{\rm d} \approx 0.35$

| $\sigma_8$ | 0.34 | 0.41 | 0.49 | 0.55 | 0.62 | 0.69 |
|---|---|---|---|---|---|---|
| percentage | 10% | 32% | 53% | 71% | 83% | 92% |

### 4.2. Abundance of arcs in rich Abell clusters

After having established what fraction of rich Abell clusters is critical to lensing, we want to calculate the abundance of arcs in these clusters. Only some fraction of critical will actually show arcs, since the background sources need to be in the right place as well. We generated the source redshift distribution as in Broadhurst, Taylor & Peacock (1995). This gives us a realistic distribution of source redshifts and magnitudes. The clusters themselves were put at the average redshift at which they are most likely to be critical to lensing, i.e. around 0.35 for the standard CDM scenario (see above). The lensed images of background sources were then calculated numerically using the thin lensing approximation (eg. Bartelmann & Weiss 1994). Lens amplification is taken into account. An example of such a calculation is shown in Figure 3 (for cluster model 41 and $\sigma_8 = 0.62$). The image is somewhat untypical in the sense that there is a more than average number of arcs visible.

### 4.3. Abundance of arcs in X-ray luminous clusters

A complete observed sample that was searched for arcs only exists for X-ray luminous clusters, that is, clusters with an X-ray luminosity $L_X(0.3-3.5\text{keV}) > 4 \times 10^{44}$ erg s$^{-1}$ (Le Fèvre et al. 1994). For this sample, with 16 clusters in the redshift range $0.2 < z < 0.6$ and an average redshift of 0.35, 11 arcs were found. The $L_X$ lower limit roughly corresponds to a limit of 1100 km s$^{-1}$ in line-of-sight velocity dispersion (Quintana & Melnick 1982). We used this to select a cluster subsample from our model catalogue that should be comparable to this observed sample. Each of these clusters was again put at $z = 0.35$, and five different backgrounds were used for each of these to assemble a sample of model images. These were searched for arcs, and the results for several $\sigma_8$ are shown in Table 2. Note that the number of clusters that are actually in the sample (i.e. that have an l.o.s. velocity dispersion of over 1100 km s$^{-1}$) increases with $\sigma_8$, since we have a fixed volume for the model catalogue and velocity dispersion increases with $\sigma_8$. A comparison between the abundances for the models to the 11/16 abundance of arcs in the observed sample of Le Fèvre et al. (1994) shows that $\sigma_8$ should be at least 0.6, but not be much larger than 0.7. Please note that these results are still preliminary, because we still have to take into account shapes and colours of the background galaxies, seeing effects in the simulated images, and we have to do the analysis at a higher resolution (i.e. comparable to the observed resolution).

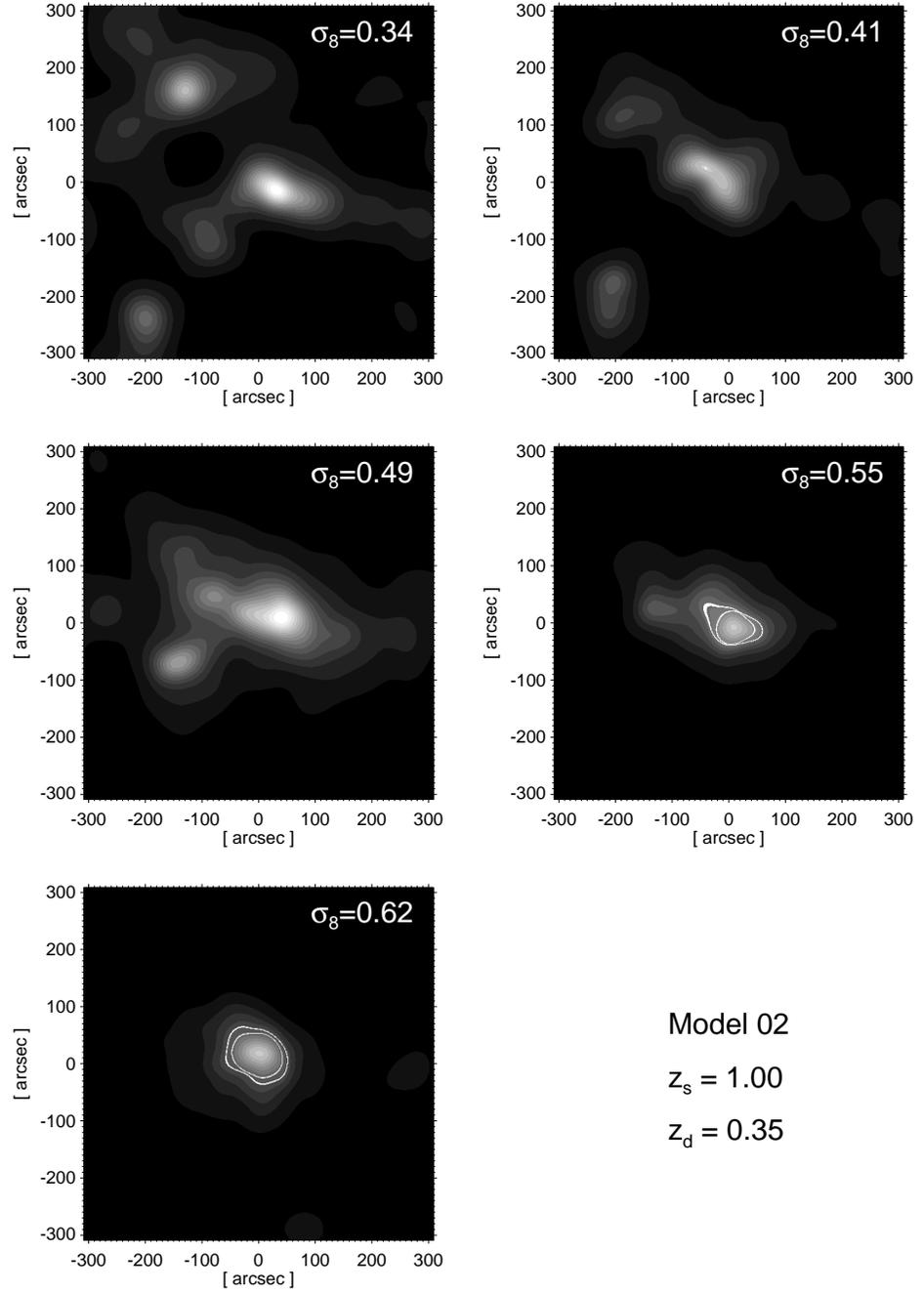

Figure 2. Surface mass density as a function of $\sigma_8$ for one of the cluster models from the catalogue. Critical curves for $z_d = 0.35$ and $z_s = 1$ are overlayed, if present. This particular model becomes critical to strong lensing for $\sigma_8 > 0.5$.

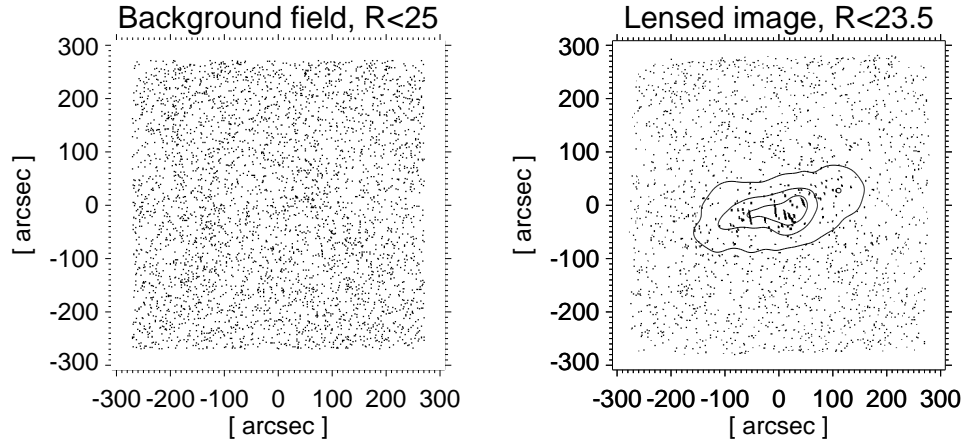

Figure 3. Example of a background source distribution for $R < 25$ (left) and its lensed image due to an intervening X-ray luminous cluster for $R < 23.5$ (right). Surface mass density contours are also shown.

Table 2. Abundance of arcs for high-$L_X$ clusters

| $\sigma_8$ | 0.49 | 0.55 | 0.62 | 0.69 |
|---|---|---|---|---|
| arcs per cluster | none | 0.4 | 0.5 | 0.8 |
| no. of clusters | 1 | 4 | 7 | 15 |

## 5. Conclusions and discussion

The (preliminary) constraint on $\sigma_8$ for standard CDM from gravitational lensing statistics, $0.6 < \sigma_8 < 0.7$, is not consistent with that obtained from cluster statistics, $0.4 < \sigma_8 < 0.5$. Note that the measurements of *COBE* imply that $\sigma_8 \approx 1.3$ for our chosen scenario (Bunn, Scott & White 1995). This is far off from all values we found here. However, the scales measured by *COBE* fall well outside the dynamical range of our simulations. So an initial fluctuation spectrum that looks like standard CDM on cluster scales, but has more power on the much larger scales that are traced by *COBE* , might fix this. But it seems more difficult for standard CDM to resolve the inconsistent values for $\sigma_8$ for the four measures we considered, which are on the same scale and within the dynamical range of the simulations. However, the discrepancy is not horrific, so a variation on CDM (including the right amount of power on large scales) might still be viable, which is what we will investigate in the near future.

**Acknowledgments.** The author acknowledges Andy Taylor for supplying the background galaxy distribution, and an European Community HCM Research Fellowship for financial support.